\documentstyle[12pt,axodraw]{article}
\begin{document}
\pagestyle{empty} \quad\vskip -1.5cm 
\centerline{\hfill  SLAC-PUB-8707}
\centerline{\hfill  hep-ph/0011163}
\centerline{\hfill  November 2000}
%%%%%%%%%%<<<<<<<<<=======
%
\vskip 1cm
\begin{center}{\bf \large
Challenges in Hyperon Decays
\footnote{Contribution to Workshop On Low-Energy Pbar Storage Ring (Pbar2000) 
3-5 Aug 2000, Chicago, Illinois}
\footnote{Work supported by the Department of Energy, Contract 
DE-AC03-76SF00515}
}

\vskip 1cm

{\large
Darwin Chang
}\\

{\em NCTS and Physics Department, National Tsing-Hua University,\\
Hsinchu 30043, Taiwan, R.O.C.}\\
and\\
{\em Stanford Linear Accelerator Center, Stanford University, Stanford, 
     CA 94309, USA}\\

\end{center}
\date{\today}
\vskip 1.5cm

\begin{abstract}
We give an personal overview of some of the unsolved problems
related to hyperon decays.  We cover nonleptonic decays,
radiative decays and magnetic moments.  Some of the theoretical
issues are also touched upon.

\end{abstract}

\section{Introduction}

While the frontier of high energy physics marches on to higher
and higher energies, there are many unsolved puzzles left over in
low energy physics.  Hyperon decays account for many of them.
These puzzles are generally not considered serious enough to
deserve major attention. Our inability to calculate accurately
with strong interaction is usually the reason invoked to explain
why these puzzles remains so and why they may not be directly
relevant to our understanding of the fundamental physics.

Nevertheless, one should be reminded that, even if their solutions
do not require changes in the fundamental physics,
it is still utmost important to understand
the dynamics of low energy hadronic physics through the thicket of
strong interaction.  In addition, one should keep in mind
that the surprises can potentially occur that alter our
understanding at the fundamental level.  One example is the potential
new source of CP that can be revealed itself in hyperon decays.
There are of course many other corners that the new physics may
have already revealed itself in the data, and yet unrecognized by
us through our inability to quantify strong interaction.
One example that stands out is the $\Delta I = 1/2$
rule that had been repeatedly used to motivate new physics in the
literature in the past.

Since hyperon decays is a vast and hard subject, it is beyond my
ability to review most of them in detail \cite{review}.
My goal is only to recount some of them with some personal
perspective so as to help motivate more future
experiments on this subject.

\section{Problems and tools}

The various tools or models was invented to understand hadronic
dynamics without facing the strong coupling issues head on.  One
can start by taking advantage of only the symmetry property which
presumably is respected by the strong dynamics except the
potential spontaneous symmetry breaking by vacuum condensates.
One first identifies, or assumes, the low energy degrees of
freedom and their associate symmetry, and proceeds to write down
the most general low energy effective theory describing these
degrees of freedom consistent with the required symmetry.  This is
the approaches taken by chiral perturbation theory (ChPT)\cite{chpt}
and QCD sum rule\cite{qcd-sum-rule}.
These low energy effective theories typically involve
infinite numbers of terms, or couplings.  To make such theories
useful some expansion parameters and the associated cut-off scheme
is necessary to derive predictions.  For example in the simplest
ChPT, the octet of Goldstone bosons associated with chiral
symmetry breaking are identified as the low energy degrees of
freedom and using its energy and momentum as expansion parameter
an infinite series of Lagrangian can be written down according to the
chiral symmetry.  The higher dimensional operators are suppressed
by powers of $p/\Lambda_\chi$ where $\Lambda_\chi$ is the chiral
symmetry breaking scale.
Similar expansion series is employed in QCD sum rule approach.

Unfortunately, there is no guarantee that either the expansion is
convergent, or, the degrees of freedom included is sufficient.
This is especially the case when the energy involved in the
hadronic process is not limited to very small $p/\Lambda_\chi$
and can sometimes be quite close to where the high resonance
occurs.  In such case it is often impossible to justify the
approximation or to account for the data without extra inputs.
However, there is so far no reliable principles about how such
extra inputs should be invoked or deployed.  As a results, the
theory has degenerated into models that are mostly invoked to
account for particular special subsets of phenomenology.  For
example, it was well known that the lower order chiral perturbation
theory cannot account for the data on hyperon radiative decays
\cite{lach,review2} even after the heavy baryon chiral perturbation 
theory formalism is employed \cite{bos1,bos2,bos3}.  
In an attempt to resolve this puzzle, some additional 
resonances are added in the analysis recently \cite{bh}.
While the resulting "model" can account for the data (in fact it
can account for the data in more than one way), however it is not
clear how much fundamental understanding is actually gained in the
process.  
With these general comments on the theoretical difficulties,
we can go over the specific challenges.

The challenges facing the hyperon decays can be summarized in 
the following categories :\\
(1)Nonleptonic decays; \\
(2)CP violation; \\
(3)Semi-leptonic decays; \\
(4)Radiative decays; \\
(5)Magnetic moments.\\
I shall leave the topic of CP violation to German Valencia and 
topic of semi-leptonic decays to Earl C. Swallow in this workshop.

\section{Nonleptonic Decays}

The nonleptonic decays, $B \rightarrow B' \pi$, includes:\\
$\Lambda \rightarrow p \pi^-$; \,\,\,
$\Lambda \rightarrow n \pi^0$; \,\,\, \\
$\Sigma^+ \rightarrow n \pi^+$; \,\,\,
$\Sigma^+ \rightarrow p \pi^0$; \,\,\,
$\Sigma^- \rightarrow n \pi^-$; \,\,\, \\
$\Xi^0 \rightarrow \Lambda \pi^0$; \,\,\,
$\Xi^- \rightarrow \Lambda \pi^-$. \,\,\,\\
The amplitude can be written as
$$ \it{M} = \bar{u}_{B'}(p') (A + B \gamma_5 ) u_B(p) $$
where amplitudes $A$'s are parity violating, S-wave while $B$'s are
parity conserving and P-wave.
%Naively, one expects each $B$'s to be smaller than the corresponding $A$'s.

\subsection{$\Delta I = 1/2$ Rule}

Each amplitude can be further decomposed into two isospin
channels. For example, $A$ amplitude for $\Lambda \rightarrow p
\pi^-$ can be decomposed into $A(\Lambda \rightarrow p \pi^-) =
\sqrt{2} A^{1/2}_{\Lambda_-} - A^{3/2}_{\Lambda_-}$ and similarly
for $B$ amplitudes. Experimentally \cite{pdg}, the ratio of
$A^{3/2}/A^{1/2}$ for each S-wave, $A$, or P-wave, $B$, amplitude
is found to be small and of the order of that in the nonleptonic
kaon decays.  For example,
$${A^{3/2} \over A^{1/2}} =\bordermatrix{&S&P \cr
                        \Lambda&0.014&0.006 \cr
                        \Sigma&-0.017&-0.047 \cr
                        \Xi&0.034&0.023 \cr}$$
The fact that these ratios, as well as the corresponding one in
kaon nonleptonic decays are all of the order of 0.02 in absolute value
indicates that $\Delta I =1/2$ rule may reflect something general
about the strangeness changing nonleptonic decays.
However, so far we have not been able to understand this selection
rule in general.

For example, in kaon decays, the factor of $20$ in $\Delta I =1/2$
enhancement is partially accounted for (by a factor of 3 to 4) by
the renormalization group (RG) effect of QCD between the weak
scale and the low energy kaon scale.  However, numerically this
enhancement is still insufficient and this actually leads to many
proposal of new physics in the literature in order to account for
it.

For hyperon, the understanding improves a little bit.  The
similar RG enhancement effect is still operating.  Pati/Woo and
Miura/Minamikawa \cite{pati} discovered another potential
enhancement effect. They observed that, using Fierz
transformation, the $O^{\Delta = 3/2}$ operator that can mediate
hyperon decays, can be written in such a way that it is symmetric
under the exchange of color indices of either $(u, s)$ pair or
$(d, s)$ pair. If the baryon are taken naively as consisting of
three valence quarks only, then the anti-symmetry of color
wavefunction as well as this color property of the $\Delta I
=3/2$ operator imply that only the  $\Delta I =1/2$ operator can
mediate the transition $<B' | \it{H}_W | B>$ where  $\it{H}_W$ is
the weak Hamiltonian, that is, $<B' | \it{O}^{\Delta =3/2} | B> =
0$. Of course, $B \rightarrow B'$ matrix elements are not the only
mechanism for $B \rightarrow B' \pi$ decays.  And, baryon
wavefunction is not exactly just three quark only.  However, it
is a good indication that the $\Delta I = 1/2$ amplitudes are
dynamically favored.  The most unsatisfactory aspect of this
explanation is probably that the same mechanism clearly does not
apply to kaons.  Therefore it does not provide a general
understanding of the strangeness changing nonleptonic decays
together.  Nevertheless, with such qualitative mechanism at hand,
one can conclude that, overall, $\Delta I = 1/2$ rule is probably
not as serious a problem theoretically as in the case of kaons.

\subsection{Nonleptonic enhancement}

The data clearly shows the typical non-leptonic branching ratios, say,
$\Gamma(\Lambda \rightarrow p \pi^-)$ is about $10^3$ larger than the
typical semi-leptonic ones, say,
$\Gamma(\Lambda \rightarrow p e \bar{\nu}_e)$.  Why?
We have not been able to account for that based on the fundamental principle.
There were various attempts to resolve this difference using quark
model plus some QCD inputs many years ago.
However, none can be considered compelling.
This problem has not received too much recent attention.
Nevertheless, I am wondering whether one can derive some better
qualitative understanding by learning from the recent works on related
issues in the charm and especially the B systems, such as heavy quark
perturbation theory \cite{hbcpt} or QCD inspired factorization models \cite{hnli}.

\subsection{S-Wave/P-Wave problem}

In chiral models or current algebra approach, S-wave ($A$) is due
to a contact term or, in ChPT language, the lowest order term in
chiral Lagrangian. It is typically represented as in Fig. 1a.
Note that the same vertex does not contribute to the P-wave
amplitudes, B.  To get P-wave amplitudes, one need to invoke the
pole diagrams in Fig. 1b and Fig. 1c.  The lowest order weak
Lagrangian gives
$$ \it{L}^0_W = h_D Tr(\bar{B} \{\lambda, B \}) + h_F Tr(\bar{B}[\lambda, B])
$$
where $\lambda = \xi^\dagger \lambda_6 \xi$, $\xi^2 = \Sigma$ is
the usual Goldstone bosons in matrix form.
The $\it{L}^0_W$ contributes to S-wave amplitudes, $A$, directly,
but not to $B$.  The two parameters $h_D$ and $h_F$ can be used
to fit the S-wave amplitudes quite well at this order. In particular,
$A(\Sigma^+ \rightarrow n \pi^+)$ is found to be zero at this
order while experimentally, indeed, it is much smaller than
others.

%%%%%%%%%%%%%%%%%%%%%%%%%%%%%%%%%%%%%%%%%%%%%%%%%%%%%%%%%%%%%%%%%%%
\begin{center}
\begin{picture}(340,100)(0,0)

 \ArrowLine(0,50)(50,50)   \Text(20,60)[c]{$B$}
 \ArrowLine(50,50)(80,20)  \Text(90,30)[c]{$B'$}
 \DashArrowLine(50,50)(80,80){5} \Text(70,80)[c]{$\pi$}
 \GBoxc(50,50)(5,5){0}
 \Text(40,10)[c]{$(a)$}

 \ArrowLine(100,50)(140,50)   \Text(120,60)[c]{$B$}
 \ArrowLine(140,50)(180,50)  \Text(160,60)[c]{$B"$}
 \ArrowLine(180,50)(210,20) \Text(215,30)[c]{$B'$}
 \DashArrowLine(180,50)(210,80){5} \Text(190,80)[c]{$\pi$}
 \GBoxc(140,50)(5,5){0}
 \GCirc(180,50){3}{0}
 \Text(150,10)[c]{$(b)$}

 \ArrowLine(230,50)(270,50)   \Text(250,60)[c]{$B$}
 \ArrowLine(270,50)(300,35)  \Text(280,35)[c]{$B"$}
 \ArrowLine(300,35)(330,20)  \Text(320,35)[c]{$B'$}
 \DashArrowLine(270,50)(300,80){5} \Text(280,80)[c]{$\pi$}
 \GBoxc(300,35)(5,5){0}
 \GCirc(270,50){3}{0}
 \Text(260,10)[c]{$(c)$}

 \end{picture}
 \\
Fig.1. Here circle and square represent strong and weak vertex
respectively.
\end{center}
%%%%%%%%%%%%%%%%%%%%%%%%%%%%%%%%%%%%%%%%%%%%%%%%%%%%%%%%%%%%%%%%%%%%

Using the fitted $h_D$ and $h_F$, together with strong
interaction parameters, one can calculate the pole diagrams and
predict the P-wave amplitudes, $B$. The result turns out to be a
poor fit of many of the $B$ amplitudes extracted from the data.
Actually, alternatively, one can also chose to use $h_D$, $h_F$
to fit the P-wave amplitudes, and use them to predict S-wave
amplitudes.  The result is an equally poor fit. The is the famous
S-wave/P-wave problem.

There are a few potential ways out of the puzzle.  One can see if
the inclusion of higher order effective Lagrangian can rectify
the situation.  However, there are just too many parameters in
the next order Lagrangian and there is no unique, convincing fit
to the data. Alternatively, one can adopt a model that include
new resonances, such as Spin 1/2 Parity-odd resonances or the
Spin 1/2 parity-even Roper resonances, as the intermediate
states, $B''$ in the pole diagrams\cite{bh}.
The result can fit both S-
and P-wave amplitudes, however the fit is clearly model dependent.

\section{Radiative Decays}

The radiative decays are :\\
$\Sigma^+(uus) \rightarrow p(uud)
\gamma$ \,\,\,
BF $\sim (1.23 \pm 0.05)\times 10^{-3}$ ; ** \\
$\Sigma^0 \rightarrow n \gamma$ \,\,\,
BF too small to see due to EM decay
$\Sigma^0 \rightarrow \Lambda \gamma$ \\
$\Xi^0 \rightarrow \Sigma^0 \gamma$; \,\,\,
BF $\sim (3.5 \pm 0.4)\times 10^{-3}$ ;  \\
$\Xi^0 \rightarrow \Lambda \gamma$; \,\,\,
BF $\sim (1.06 \pm 0.16)\times 10^{-3}$ ;  \\
$\Lambda \rightarrow n \gamma$; \,\,\,
BF $\sim (1.75 \pm 0.15)\times 10^{-3}$ ;  \\
$\Xi^- \rightarrow \Sigma^- \gamma$; \,\,\,
BF $\sim (1.27 \pm 0.23)\times 10^{-4}$ ; ** \\
$\Sigma^0 \rightarrow \Lambda \gamma$ \,\,\,
BF $\sim 100\% $ electromagnetic decay .  \\
(Note that ** are charged modes between U-spin doublets.)

Generally, the amplitudes can be rewritten as
$$
A = - {e \over M_B + M_{B'}} i \epsilon^{*\mu} q^\nu \bar{u}(p')
\sigma_{\mu \nu} (C + D\gamma_5) u(p)
$$
where $C$ is parity conserving magnetic dipole transition (M1),
$D$ is parity violating electric dipole transition (E1). The decay
rate is $\Gamma \propto |C|^2 + |D|^2$.  The asymmetry is $A_\gamma =
{2 Re(C^* D) \over |C|^2 + |D|^2}$, that is, one needs both
nonzero $C$ and $D$ amplitudes to get nonzero asymmetry.  There
is an old theorem\cite{hara} by Hara regarding the vanishing of $D$
amplitudes.

\subsection{Hara's Theorem}

\underline{Hara's Theorem}:
Assuming that the amplitudes are not singular in the $SU(3)_F$
limit, parity violating $D$ amplitudes must vanish for decays between
states of a $U$-spin doublet in the $SU(3)_F$ limit.\\
The theorem implies that
$$D(\Sigma^+ \rightarrow p  \gamma) = D(\Xi^- \rightarrow \Sigma^- \gamma) = 0
$$
and as a result the asymmetries
$A_\gamma(\Sigma^+ \rightarrow p
\gamma) = A_\gamma(\Xi^- \rightarrow \Sigma^- \gamma) = 0 $

Accepting the assumption of Hara's theorem, even when $SU(3)$ breaking is taken
into account, asymmetry $A_\gamma$ should remain small.
Unfortunately, experimentally $A_\gamma(\Sigma^+ \rightarrow p  \gamma)$,
which is the only one measured to be nonzero,
was found to be $-0.76 \pm 0.08$, indeed large and negative.
The others, given in Table 1., have much larger error.

The hadronic models did not have a great deal of success in explaining
the details of the data.  All the models of this type (except those that
include vector mesons) preserve Hara's theorem in their
formulations.  General analysis which include $SU(3)$
breaking\cite{vasa} actually predicts a small and positive asymmetry
for the $\Sigma^+$ decay while the data is negative and relatively large.
Models that assume vector-meson dominance
\cite{vector} can introduce effects that violate Hara's theorem due to
mixing of the vector meson with the photon.  In models using quarks,
it was pointed out \cite{kamal} that the diagrams in which a $W$-boson
is exchanged between two constituent quarks can give rise to violation
of Hara's theorem.  In addition, models which include vector-meson
dominance are in better agreement with the data, though the situation
is still not satisfactory.  Among hadronic models, the observed negative
asymmetry parameter for $\Sigma^+$ decay is best accounted for using
QCD sum-rules \cite{qcd-sum-rule}.  Other approaches can be found in
Refs.~\cite{other,neuf,jenk}.  A detailed overview on both
experimental and theoretical aspects of weak radiative decays of
hyperons is given in Refs.~\cite{timm,lach,review2}.

Many papers questioned the assumption of Hara Theorem based on
general ground\cite{antihara,antihara2}, including some recent
references\cite{antihara2}. The point is that it is possible that
the parity violating amplitude, $D$, becomes singular when one
takes the $SU(3)$ limit.  It reflects that how to treat the
$SU(3)$ breaking can have a crucial influence on the outcome of
the analysis.  As elaborated in the magnetic moment section
later, we believe that more careful analyses are still wanting in
this aspect in the literature.

One may try to employ ChPT, that was useful in describing low
energies hadronic processes involving only mesons, to tackle
hyperon.  For application to processes involving baryons, it is
most consistently formulated in the heavy-baryon formulation
\cite{hbcpt}, in which the $SU(3)$ invariant baryon mass,
$\dot{m}$, is removed by a field transformation (see
Ref.~\cite{bos1} for detail). In this approach an amplitude for a
given process is expanded in external pion four-momenta, $q$,
baryon {\em residual\/} four momenta, $k$, and the quark mass,
$m_s$.  If one neglects the up and down quark mass.  One can
collectively write down $q$, $k$, and $m_s$ as $E$ ( and adopt
the convention that $k$ and $m_s$ are of the same order in the
chiral expansion). The perturbation theory is reliable only when
$E$ is smaller than the chiral symmetry breaking scale
$\Lambda_{\chi}$.  In the heavy-baryon formulation there is an
additional expansion in $1/\dot{m}$.  However, all these terms
can be effectively absorbed in counterterms of the theory
\cite{bos1,bos2,bos3}.

Weak radiative decays of hyperons have been studied before in the
context of ChPT by Jenkins, Luke, Manohar and Savage \cite{jenk}
and Neufeld \cite{neuf}.  Jenkins {\it et al.} and Neufeld
calculated the amplitude up to the one-loop level.  These loop
diagrams give contributions to the amplitudes which are at least
of order $O(E^2)$ in the chiral expansion.  However, tree-level
direct emission diagrams from the next-to-leading order weak
Lagrangian, which give contribution of order $O(E)$ to the
amplitudes, were not considered\cite{bos1,bos2}. In a series of
paper\cite{bos1,bos2,bos3} we consistently calculated the
leading-order amplitude of weak radiative decays of hyperons in
ChPT.  At this order, no loop contribution need to be considered.
However, one does need to take into account the higher-order
terms in the weak chiral Lagrangian.  We showed that it gives rise
to violation of Hara's theorem.  As a consequence the decay rates
for the charged decays $\Sigma^+\rightarrow p+\gamma$ and $\Xi^-
\rightarrow \Sigma^- \gamma$ can be accounted for consistently.
We showed that, in leading order, ChPT predicts the ratios of the
decay amplitudes of all the neutral channels as functions of the
baryon masses only.  We compared these predictions with the data.
Furthermore, the asymmetry parameters still vanish in this
leading order calculation.  However, this is not necessarily
inconsistent with the data in the expansion scheme of ChPT.

In particular, the diagrams contributing to the leading-order
amplitude are tree diagrams given in Fig.~2.  There are two kinds
of diagrams: the direct emission diagrams, Fig.~2a, and the baryon
pole-diagrams, Fig.~2b and Fig.~2c.  Loop diagrams give rise to
contributions of higher order.  The pole-diagrams only contribute
to the parity-conserving form factor $C$, in accordance with the
Lee-Swift theorem \cite{lee-swift}. Unfortunately as concluded in
Ref.\cite{bos3}, within the context of simplest heavy baryon
chiral perturbation theory without inclusion of any additional
resonances, the understanding of many of the data in radiative
hyperon decays, including the asymmetry, is still beyond reach.
However, if one is willing to introduce additional resonances,
together with the additional parameters that come with it, it is
possible to fit the data without some models\cite{bh}.
%%%%%%%%%%%%%%%%%%%%%%%%%%%%%%%%%%%%%%%%%%%%%%%%%%%%%%%%%%%%%
\begin{center}
\begin{picture}(340,100)(0,0)

 \ArrowLine(0,50)(50,50)   \Text(20,60)[c]{$B$}
 \ArrowLine(50,50)(80,20)  \Text(90,30)[c]{$B'$}
 \Photon(50,50)(80,80){3}{5} \Text(70,80)[c]{$\gamma$}
 \GBoxc(50,50)(5,5){0}
 \Text(40,10)[c]{$(a)$}

 \ArrowLine(100,50)(140,50)   \Text(120,60)[c]{$B$}
 \ArrowLine(140,50)(180,50)  \Text(160,60)[c]{$B"$}
 \ArrowLine(180,50)(210,20) \Text(215,30)[c]{$B'$}
 \Photon(180,50)(210,80){3}{5} \Text(190,80)[c]{$\gamma$}
 \GBoxc(140,50)(5,5){0}
 \GCirc(180,50){3}{0}
 \Text(150,10)[c]{$(b)$}

 \ArrowLine(230,50)(270,50)   \Text(250,60)[c]{$B$}
 \ArrowLine(270,50)(300,35)  \Text(280,35)[c]{$B"$}
 \ArrowLine(300,35)(330,20)  \Text(320,35)[c]{$B'$}
 \Photon(270,50)(300,80){3}{5} \Text(280,80)[c]{$\gamma$}
 \GBoxc(300,35)(5,5){0}
 \GCirc(270,50){3}{0}
 \Text(260,10)[c]{$(c)$}

 \end{picture}
 \\
Fig.2. Here circle and square represent strong and weak vertex
respectively.
\end{center}
%%%%%%%%%%%%%%%%%%%%%%%%%%%%%%%%%%%%%%%%%%%%%%%%%%%%%%%%%%%%%%%%%%%%%%%%%%

\section{Magnetic Moments}

The magnetic moments of the octet baryons were found to obey
approximate $SU(3)$ symmetry a long time ago by Coleman and
Glashow (CG)\cite{coleman}.  In the $SU(3)$ symmetric limit, the
nine observable moments (including the transitional moment between
$\Sigma^0$ and $\Lambda$) can be parameterized in terms of two
parameters, and as a result obey approximate relationships.  The
two parameter result of CG can in fact fit the observed magnetic
moments up to about the $20 \%$ level.  However, since at present
the moments have been measured with an accuracy of better than $1
\%$ \cite{pdg}, an improved theoretical understanding is clearly
desirable.

Many attempts were made trying to improve the numerical
predictions of CG by including the $SU(3)$ breaking effects using
ChPT \cite{caldi,krause,jenk2,luty}. However, many of these
efforts resulted in numerical fits {\em worse\/} than the leading
order $SU(3)$ invariant one by CG. For example, Caldi and Pagels
\cite{caldi} found that the leading $SU(3)$ breaking corrections,
in their scheme for ChPT, appear in the non-analytic forms of
$\sqrt{m_s}$ and $m_s \ln m_s$.  They showed that the
$\sqrt{m_s}$ corrections are in fact at least as large as the
$SU(3)$ invariant zeroth order terms, which casts doubt on the
applicability of ChPT.  Caldi and Pagels suggested that this
``failure'' of ChPT might be attributed to the large mass of kaon
in the loops and the fact that the leading correction is of
non-analytic form.  Such non-analytic contributions were indeed
pointed out earlier by Li and Pagels \cite{li} and others
\cite{langacker}, however, the non-analyticity appears only in
the $SU(3)$ invariant chiral symmetry breaking mass, not in
$SU(3)$ breaking parameters.  More recently, similar large
corrections to the baryon magnetic moments non-analytic in $m_s$
have been found, by calculating them up to the one-loop level in
ChPT \cite{krause,jenk,luty}. By only using the $\sqrt{m_s}$
terms Jenkins {\em et al.} \cite{jenk} could improve the accuracy
of the Coleman-Glashow results from 20 \% to about 10 \%.
However, this could only be achieved by using in kaon loops a
{\em different\/} value of the meson decay constant than in pion
loops, with the effect that the magnitude of the kaon loops is
artificially reduced. In addition, Krause \cite{krause} showed
that $m_s \ln m_s$ corrections are just as important, which
disagrees with Refs.~\cite{jenk,luty}. Also, Krause further
argued that the non-analytic contributions are not a good
approximation of the loop integrals at all.

\subsection{Okubo relation}

Shortly after CG, Okubo \cite{okubo} derived a relation
$$6\mu_{\Lambda} + \mu_{\Sigma^-} - 4\sqrt{3} \mu_{\Lambda\Sigma^0}
  - 4\mu_n + \mu_{\Sigma^+} - 4 \mu_{\Xi^0} = 0,
$$
where $\mu_{\Lambda\Sigma^0}$ is the $\Sigma^0\rightarrow\Lambda$
transition moment.  This relation can be obtained if one assumes
that $SU(3)$ breaking corrections to the moments are linear in
the quark mass matrix $\sigma$, defined by $\sigma \equiv {\rm
diag}(0,0,m_s)$. The $SU(3)$ breaking terms introduce additional
5 parameters. The resulting 7 parameter prediction can fit the
current 9 high precision observables to within 1.5\% \cite{bos4}.

Within the context of ChPT, there is no unique way of treating the
$SU(3)$ breaking yet.  However, the nice fit of Okubo relation
can be taken as a hint. In Ref.\cite{bos5}, it was shown that,
even with in the scheme of ChPT used in Ref.\cite{jenk2}, while
the one-loop non-analytic contributions of the form $\sqrt{m_s}$
satisfies the Okubo relation, the nonanalytic contributions of
the form $m_s \ln m_s$ does not! In Ref.\cite{bos4}, a
formulation of ChPT was suggested such that the leading $SU(3)$
breaking correction is indeed linear in $m_s$.

%\section{Issues}
%\subsection{A subheading}

\section{Summary}

We had shown that there are still many issues that we do not
understand associated with hyperon decays based only on
fundamental principles. It is possible that some of the
difficulties are such that, due to our inability of dealing with
strong interaction, we cannot understand them without invoking
some phenomenological models.  It is also possible that some of
the current data that makes it so hard for us to understand may
be changed by improved measurement in the future.

This work was supported in parts by National Science Council of R.O.C., 
and by U.S. Department of Energy.    
We wishes to thank SLAC Theory Group for hospitality.
%

%%%%%%%%%%%%%%%%%%%%%%%%%%%%%%%%%%%%%%%%%%%%%%%%%%%%%%%%%%%%%%%%%
%
%   BEGIN OF TABLES
%
%%%%%%%%%%%%%%%%%%%%%%%%%%%%%%%%%%%%%%%%%%%%%%%%%%%%%%%%%%%%%%%%%

\begin{table}[t]
\caption{Present status of decay rates and asymmetry parameters.  The
numbers are the combined weighted mean from Ref.~\protect\cite{lach}.
Both the decay rate and the asymmetry parameter for
$\Sigma^0\rightarrow\Lambda+\gamma$ have not been measured.}
\label{data}
\begin{tabular}{cccc}
$B_i\rightarrow B_f+\gamma$
& $\Gamma$ [$10^{-18}$ GeV]
%& $|A|^2 + |B|^2$ [GeV$^{-2}$]
& $\alpha$
& Ref.
\\
\hline
$\Lambda\rightarrow n+\gamma$
& $4.07 \pm 0.35$
%& $(3.0 \pm 0.26) \times 10^{-15}$
& --
& \cite{lam}
\\
$\Xi^0\rightarrow\Lambda+\gamma$
& $2.4 \pm 0.36$
%& $(1.21 \pm 0.18 )\times 10^{-15}$
& $0.43 \pm 0.44$
& \cite{xinlam}
\\
$\Xi^0\rightarrow\Sigma^0+\gamma$
& $8.1 \pm 1.0$
%& $(1.60 \pm 0.19) \times 10^{-14}$
& $0.20 \pm 0.32 $
& \cite{xinsin}
\\
$\Sigma^+\rightarrow p+\gamma$
& $10.1 \pm 0.5$
%& $(2.8 \pm 0.14) \times 10^{-15}$
& $-0.76 \pm 0.08 $
& \cite{timm,sip}
\\
$\Xi^-\rightarrow\Sigma^-+\gamma$
& $0.51 \pm 0.092$
%& $(9.8 \pm 1.76 )\times 10^{-16}$
& $1.0 \pm 1.3$
& \cite{xim}
\end{tabular}
\end{table}
%%%%%%%%%%%%%%%%%%%%%%%%%%%%%%%%%%%%%%%%%%%%%%%%%%%%%%%%%%%%%%%%%%%%%%%

\end{document}